# MAP Format for Representing Chemical Modifications, Annotations, and Mutations in Protein Sequences: An Extension of the FASTA Format

Akshay Shendre[1], Naman Kumar Mehta[1], Anand Singh Rathore[1], Nishant Kumar[1], Sumeet Patiyal[2], and Gajendra P. S. Raghava*[1]

1. Department of Computational Biology, Indraprastha Institute of Information Technology, Okhla Phase 3, New Delhi-110020, India.

2. Cancer Data Science Laboratory, National Cancer Institute, National Institutes of Health, Bethesda, Maryland, USA.

**Mailing Address of Authors**

Akshay Shendre (AS): akshays@iiitd.ac.in     ORCID ID: https://orcid.org/0009-0006-1239-5881

Naman Kumar Mehta (NKM): namanm@iiitd.ac.in   ORCID ID: https://orcid.org/0009-0009-0244-2826

Anand Singh Rathore (ASR): anandr@iiitd.ac.in   ORCID ID: https://orcid.org/0009-0004-8907-7174

Nishant Kumar (NK): nishantk@iiitd.ac.in     ORCID ID: https://orcid.org/0000-0001-7781-9602

Sumeet Patiyal (SP): sumeet.patiyal@nih.gov    ORCID ID: https://orcid.org/0000-0003-1358-292X

Gajendra P. S. Raghava (GPSR): raghava@iiitd.ac.in   ORCID ID: https://orcid.org/0000-0002-8902-2876

**\*Corresponding Author**

Prof. Gajendra P. S. Raghava

Indraprastha Institute of Information Technology, Delhi

Okhla Industrial Estate, Phase III (Near Govind Puri Metro Station)

New Delhi, India – 110020 Office: A-302 (R&D Block)

Phone: 011-26907444

Email: raghava@iiitd.ac.in

Website: http://webs.iiitd.edu.in/raghava/

# Abstract

Several formats, including FASTA, PIR, GenBank, EMBL, and GCG, have been developed for representing protein sequences composed of natural amino acids. Among these, FASTA remains the most widely used due to its simplicity and human readability. However, FASTA lacks the capability to represent chemically modified or non-natural residues, as well as structural annotations and mutations in protein variants. To address some of these limitations, the PEFF format was recently introduced as an extension of FASTA. Additionally, formats such as HELM and BILN have been proposed to represent amino acids and their modifications at the atomic level. Despite their advancements, these formats have not achieved widespread adoption within the bioinformatics community due to their complexity. To complement existing formats and overcome current challenges, we propose a new format called MAP (Modification and Annotation in Proteins), which enables comprehensive annotation of protein sequences. MAP introduces meta tags in the header for protein-level annotations and inline tags within the sequence for residue-level modifications. In this format, standard one-letter amino acid codes are augmented with curly-brace tags to denote various modifications, including phosphorylation, acetylation, non-natural residues, cyclization, and other residue-specific features. The header metadata also captures information such as organism, function, and sequence variants. We describe the structure, objectives, and capabilities of the MAP format and demonstrate its application in bioinformatics, particularly in the domain of protein therapeutics. The MAP format offers several advantages, including seamless integration of annotations within sequence data, enriched biological context, and compatibility with existing computational pipelines. To facilitate community adoption, we are developing a comprehensive suite of MAP-format resources, including a detailed manual, annotated datasets, and conversion tools, available at http://webs.iiitd.edu.in/raghava/maprepo/.

## Author's Biography

1. Akshay Shendre is a Project Fellow in Computational Biology at the Department of Computational Biology, Indraprastha Institute of Information Technology, New Delhi, India.
2. Naman Kumar Mehta is pursuing a Ph.D. in Computational Biology at the Department of Computational Biology, Indraprastha Institute of Information Technology, New Delhi, India.

3. Anand Singh Rathore is pursuing a Ph.D. in Computational Biology at the Department of Computational Biology, Indraprastha Institute of Information Technology, New Delhi, India.
4. Nishant Kumar is currently working as Ph.D. in Computational biology from Department of Computational Biology, Indraprastha Institute of Information Technology, New Delhi, India.
5. Sumeet Patiyal is currently working as a postdoctoral visiting fellow Cancer Data Science Laboratory, National Cancer Institute, National Institutes of Health, Bethesda, Maryland, USA.
6. Gajendra P. S. Raghava is currently working as Professor and Head of Department of Computational Biology, Indraprastha Institute of Information Technology, New Delhi, India.

## Introduction

The post-genomics era is characterized by an exponential surge in protein sequence data, primarily driven by advancements in next-generation sequencing technologies. A significant challenge arising from this data deluge is the effective management of comprehensive protein information, which encompasses not only the primary amino acid sequence but also its structure, biological function, origin, post-translational modifications, and data provenance. To address this challenge and facilitate the organization and exchange of such diverse data within the scientific community, several formats have been developed over the decades. These include Protein Information Resource (PIR) format [1], an early standard designed to provide comprehensive annotation; the GCG sequence format [2], historically associated with the Wisconsin Package and frequently utilized for its analytical capabilities; the European Molecular Biology Laboratory (EMBL) format [3], a comprehensive format that stores both nucleotide and protein [4]sequence data alongside extensive annotations; and the GenBank format [5], the widely adopted format from the National Center for Biotechnology Information (NCBI) [6], which also supports rich annotation of sequence data. Furthermore, other notable formats exist, such as the SWISS-PROT format [7], recognized for its high level of manual curation and detailed functional information. Among these, FASTA has achieved widespread adoption within the scientific community due to its inherent simplicity, human readability [4], compactness, and broad compatibility across various bioinformatics tools. However, a fundamental limitation of the FASTA format is its restriction to representing linear amino acid sequences. This limitation underscores the persistent need for enhanced formats capable of accommodating the increasing complexity of protein information.

The landscape of drug discovery is increasingly transitioning towards protein and peptide-based therapeutics as promising alternatives to traditional small-molecule approaches. This shift is motivated by the intrinsic advantages of biomolecules, including their high target specificity and reduced off-target effects. The growing importance of this therapeutic modality is evidenced by the substantial increase in the number of FDA-approved protein therapeutics, as catalogued in dedicated resources such as THPdb and THPdb2 [8,9], with the count rising from 239 in 2017 to 894 in 2023 [9]. Recognizing the inherent limitations of native proteins, such as restricted stability and suboptimal bioavailability, most approved protein and peptide drugs undergo chemical modifications. These modifications, including phosphorylation, methylation, PEGylation, cyclization, and the incorporation of non-natural amino acids, enhance resistance to proteolytic degradation and facilitate targeted delivery [10–12]. In addition, most of the databases of therapeutic molecules, like CPPsite2, CancerPDF, CAMPR4, and PEPlife, contain chemically modified peptides and proteins [13–16]. There is a need to develop a format that can manage chemically modified proteins in a human-readable format like FASTA. Though well-established chemistry formats like SMILES, InChI, and MOL are effective for representing small-molecule chemical structures, they are not suitable for managing biopolymers [17,18]. In recent years, attempts have been made to extend these concepts to accommodate chemically modified peptides. Formats such as Hierarchical Editing Language for Macromolecules (HELM) and Boehringer Ingelheim Line Notation (BILN) enable atomic-level representation of peptides, capturing complex modifications and branching architectures[19,20]. HELM utilizes a graph-based notation with a defined monomer library, allowing the representation of branched structures and chemical modifications beyond linear sequences [21,22]. Similarly, BILN is a human-readable linear format capable of representing complex peptides with multiple chains, cycles, and staples [20,23]. These advancements bridge the representational gap between small molecules and macromolecules, facilitating more accurate representation of modified peptides in drug design. However, while these formats significantly enhance expressive power, they introduce increased complexity, rendering them directly incompatible with legacy FASTA-based tools and necessitating specialized parsers and knowledge of the notation. Consequently, neither HELM nor BILN has achieved widespread adoption in general sequence databases or within the broader biology community, which often prioritizes simplicity and backward compatibility.

Within the proteomics field, an awareness of these limitations has motivated initiatives such as the Proteomics Standards Initiative's Extended FASTA Format (PEFF). PEFF augments FASTA by incorporating reserved keywords and metadata lines to encode variants and post-translational modifications (PTMs) in a machine-readable text [24,25]. Notably, PEFF is designed to maintain near backward-compatibility with FASTA, enabling existing software to disregard the additional information and still process the core sequences. This underscores a critical design principle: any enhanced format should ideally preserve the ability of legacy tools to interpret the fundamental sequence information. While a completely novel format might offer expanded features, its utility is limited if it cannot be readily integrated into common workflows or necessitates extensive conversion.

In this study, we present the Modification and Annotation in Proteins (MAP) format, a novel approach that extends FASTA in-line with rich annotation capabilities while striving to maintain human readability and compatibility with existing tools. The MAP format embeds modification markers and annotations directly into the sequence string using a simple curly brace notation and employs structured tags in the header line for protein-level metadata. The design objectives of MAP are to: (1) capture residue-level modifications (chemical modifications, non-standard amino acids, binding sites, mutations) as an integral part of the sequence; (2) include protein-level descriptors (e.g., source organism, functional class, database IDs) in a standardized manner within the header; (3) remain easily readable and writable by humans; and (4) enable straightforward conversion to classical FASTA by stripping tags, thereby ensuring that core sequence information remains accessible to legacy applications.

In developing MAP, our aim was to bridge the practical gap between extremely simple formats (like FASTA) and highly complex notations (like HELM and BILN, providing a middle-ground format that is both expressive and convenient. In the subsequent sections, we will detail the MAP format, including its syntax and supported annotation types. We will then demonstrate MAP's capabilities through illustrative example sequences and benchmark tests assessing its compatibility and data richness. Finally, we will compare MAP to FASTA, HELM, and BILN, discussing the advantages and potential limitations of our approach. By introducing the MAP format, we aim to facilitate improved integration of sequence modifications into mainstream bioinformatics workflows, enabling the generation of richer protein datasets and more comprehensive analyses.

## Methods

**MAP Format Design and Syntax**

**Overall structure:** The structure of a MAP format entry closely mirrors that of a FASTA entry, comprising a header line followed by one or more sequence lines. Consistent with FASTA, the header line initiates with the > character and contains an identifier, optionally succeeded by one or more metadata tags enclosed in curly braces. The subsequent sequence lines contain the amino acid sequence represented by standard one-letter codes, with the key distinction of embedded curly-brace tags immediately following specific residues to indicate modifications or annotations. This design strategy ensures that a MAP entry maintains a visual and structural resemblance to FASTA, thereby facilitating its adoption within the scientific community. Bioinformatics tools not designed to interpret MAP will typically process the header up to the first whitespace character, effectively capturing the identifier, and may disregard any subsequent brace-enclosed text as unfamiliar. Similarly, within the sequence, if a parser encounters brace characters, they may be skipped or treated as unrecognized characters. Critically, the order of the underlying amino acid letters remains intact. Consequently, the removal of {...} tags and replacing the {nnr: …} non-natural residue tags with 'X', from a MAP entry, yields a valid standard FASTA sequence containing the fundamental amino acid sequence. This inherent simplicity in design confers a significant degree of backward compatibility to MAP – the core sequence information is preserved and accessible to any tool lacking the capacity to interpret the annotations. Figure 1 shows the overall architecture of the MAP format.

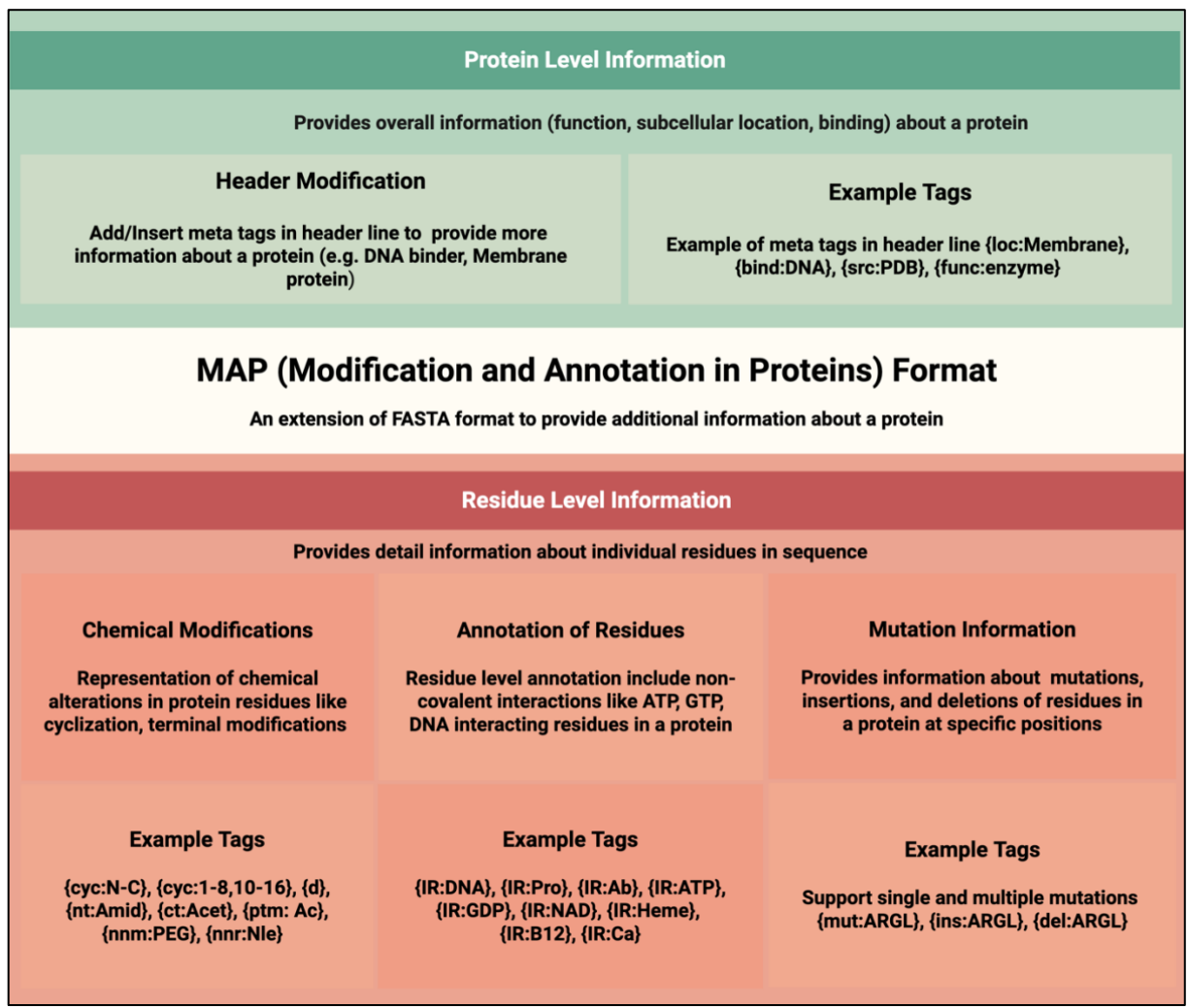

**Figure 1: Overall architecture of the MAP format**

**Header metadata tags:** Subsequent to the identifier in the header line, MAP facilitates the inclusion of structured tags in the format {key:value} to encode protein-level information. Each tag is encapsulated within its own set of braces, and multiple tags are delimited by spaces. Table 1 provides a summary of exemplary header tags that can be defined within the MAP format. These tags offer a controlled and standardized approach to incorporate information that is typically represented as free text in FASTA headers. Furthermore, users retain the flexibility to define custom tags according to their specific requirements (e.g., {disease:Cancer} or {isoform:2}). By structuring metadata in this manner, MAP enables more efficient programmatic parsing of protein attributes directly from the header line, thereby enhancing data integration capabilities. Importantly, the presence of these tags does not impede the functionality of standard FASTA parsers in reading the primary identifier. Any well-formed parser that terminates at the first whitespace character or recognizes the > symbol as the start of a new entry will still successfully capture the main identifier (i.e., everything preceding the first {}). Consequently, these metadata tags can be safely ignored by legacy software while simultaneously being leveraged by MAP-aware tools for advanced querying functionalities (e.g., retrieving all sequences originating from *Homo sapiens* or all membrane proteins annotated with {loc:Membrane}).

**Table 1:** Shows example tags that can be inserted in the header line of a protein/peptide, as well as a description of each tag.

| Tag | Description with example |
|---|---|
| {id:} | Accession or unique identifier. Example: {id:P99999} |
| {name:} | Descriptive protein name or gene name. Example: {name:Cytochrome c} |
| {org:} | Organism of origin (typically Latin binomial). Example: {org:*Homo sapiens*} |
| {func:} | Function or role of the protein. Example: {func:Electron transport} |
| {loc:} | Subcellular localization. Example: {loc:Mitochondrion} |
| {src:} | Data source or database. Example: {src:UniProt} |
| {len:} | Sequence length (number of amino acids). Example: {len:104} |
| {exp:} | Experimental evidence/status. Example: {exp:Validated} |
| {bind:} | Known binding target class. Example: {bind:DNA} for a DNA-binding |
| {target:} | If the protein is a target of a drug or process (e.g., {target:drug} |

**Residue-level annotation tags:** The core innovation of MAP lies in how it represents modifications and annotations within the sequence. MAP uses a simple convention: any residue that has an associated modification or special annotation is immediately followed by a tag in curly braces that describes that feature. The tags have human-readable prefixes indicating the type of annotation, followed by a code or value. Importantly, the presence of the tag does not break the sequence flow – e.g., ACDE{ptm:Phos}FG is interpreted as a sequence of amino acids A, C, D, E, F, G, where residue E is annotated as phosphorylated. The braces and their content are effectively out-of-band information attached to that residue. Table 2 shows examples of residue-level chemical modifications for each type of modification.

**Table 2: Shows different types of chemical modifications in a sequence, their tags, descriptions, and examples.**

| Tag | Description | Example sequences |
|---|---|---|
| **Cyclic Residues** | | |
| {cyc:N-C} | Standard head-to-tail cyclization. | ACDAFIHIK{cyc:N-C} |
| {cyc:2-4} | Single disulfide bridges between the 3$^{rd}$ and 4$^{th}$ Cysteine | ACDCFGHIK{cyc:2-4} |
| {cyc:1-3,4-6} | Multiple disulfide bridges | CICKCGC{cyc:1-3,4-6} |
| {cyc} | Undefined or non-standard cyclization | ACDECCHIKC{cyc} |
| **Terminal Modification** | | |
| {nt:Amid} | Amide group at the N-terminal. | ACDIFA{nt:Amid} |
| {ct:Acet} | Addition of an acetyl group at the C-terminal. | LMNIAC{ct:Acet} |
| **D-Amino Acids** | | |
| A{d} | Representation of D-Alanine. | GA{d}CDEFGH |
| F{d} | Representation of D-Phenylalanine. | AKDEF{d}GHCK |
| **Post-translational Modification** | | |
| {ptm:Glyc} | Glycan group at the prefix residue | GRPN{ptm:Gly}QR |
| {ptm:Ac} | Acetyl group at the prefix residue | GGLK{ptm:Ac}MN |
| **Non-natural Modification** | | |
| {nnm:PEG} | Polyethylene glycol | AGRK{nnm:PEG}DE |
| {nnm:Fluoro} | Fluorine atoms | STY{nnm:Fluoro}FGHIK |
| **Non-natural residues** | | |
| {nnr:Nle} | Methionine analog | LN{nnr:Nle}ASD |
| {nnr:Hph} | Extended Phenylalanine analog | LF{nnr:Hph}HL |
| **Conjugation of macromolecules** | | |
| {conj:Lipid} | Lipid molecules. | G{conj:Lipid}HIK |
| **Isotopic & Fluorescent labeling** | | |
| {iso:13C} | 13C isotope. | CA{iso:13C}KIDC |

All modification annotation tags share the general format {prefix:Code} (or {prefix} if no additional code is needed). The letter immediately preceding an opening brace is the residue that the tag applies to, except in case of non-natural residue with tags {nnr:Name}, in case of which the tags itself define the complete residue. In some cases, tags at the sequence termini (for modifications of the N- or C-terminus) are placed at the very end of the sequence line. Below, we outline the categories of residue-level tags supported in MAP and their usage:

- **Cyclization tags (cyc):** Cyclization, the formation of covalent linkages resulting in a ring structure within a peptide, is denoted in MAP using the {cyc:...} tag. For standard head-to-tail cyclization (N-terminal α-amino to C-terminal carboxyl), the tag {cyc:N-C} is appended to the end of the sequence, signifying a cyclic peptide. Internal cyclizations, such as disulfide bonds between cysteine residues, are represented numerically: {cyc:X-Y} indicates a linkage between the residues at positions X and Y in the sequence. For instance, the sequence ACDCFGHIK{cyc:2-4} denotes a disulfide bond between the cysteines at positions 2 and 4. Multiple cyclizations can be specified by comma-separated pairs (e.g., {cyc:1-8,2-5} for two disulfide bonds within a peptide).
- **Terminal modifications (nt and ct):** Modifications occurring at the N-terminus or C-terminus of peptides, such as N-terminal acetylation, pyroglutamate formation, or C-terminal amidation, are indicated in MAP using {nt:Type} for N-terminal and {ct:Type} for C-terminal modifications. These tags are placed at the end of the sequence line but conceptually apply to the respective terminus. For example, ACDEFGHKL{nt:Acet}{ct:Amid} represents the sequence "ACDEFGHKL" with an acetyl group at the N-terminus and an amidated C-terminus. Common codes include Acet (acetylation), Amid (amidation), Formyl (formylation), Glyco (glycosylation), Me (methylation), and others. An undefined modification can be denoted by {nt} or {ct} alone.
- **D-amino acids (d):** While standard proteins are composed of L-amino acids, many bioactive peptides incorporate D-amino acids, which are enantiomers of their L-counterparts. A D-residue possesses the same mass and composition as its L-isomer but exhibits different chirality, potentially significantly impacting peptide stability and function. In the MAP, a D-amino acid is indicated by appending {d} immediately after the one-letter amino acid code (e.g., A{d} signifies a D-alanine residue). A peptide entirely composed of D-amino acids would have each residue followed by {d} (e.g., A{d}R{d}G{d} represents D-Ala-D-Arg-D-Gly). The {d} tag requires no further specification as it solely denotes the change in chirality. This straightforward notation allows MAP to represent peptides with mixed chirality, a distinction not typically captured by FASTA, which uses the same letter for both enantiomers. For example, GA{d}CDEF{d}GHIK indicates a peptide with D-alanine at position 2 and D-phenylalanine at position 6.
- **Post-translational modifications (ptm):** Post-translational modifications (PTMs) are covalent modifications occurring on proteins after translation, including phosphorylation, glycosylation, and methylation [26690490]. MAP utilizes the prefix ptm followed by a short code to denote these modifications (e.g., {ptm:Phos} for a phosphorylated residue, typically Ser, Thr, or Tyr; {ptm:Glyc} for glycosylation, often on Asn or Ser/Thr; {ptm:Ac} for acetylation; {ptm:Me} for methylation; {ptm:Ub} for ubiquitination). The tag is placed immediately after the modified residue (e.g., S{ptm:Phos} indicates a phosphorylated serine). Multiple distinct PTMs can be present within a single sequence at different positions (e.g., MGAS{ptm:Phos}TK). A curated list of common PTM codes (e.g., Phos, Glyc, Ac, Me, Ub, Sumo, OH, Palm) is provided, and an unspecified PTM can be denoted by {ptm} alone if necessary.

- **Non-natural modifications (nnm)**: Beyond biological PTMs, proteins, particularly engineered or synthetic peptides, may undergo chemical derivatizations not typically found in nature. These are labeled using nnm (non-natural modification). Examples include PEGylation ({nnm:PEG}), biotinylation ({nnm:Biotin}), and the introduction of novel chemical groups ({nnm:Fluoro} for fluorine atom addition). The tag is placed after the modified residue (e.g., K{nnm:PEG} denotes a lysine modified by a PEG chain; C{nnm:Biotin} for a cysteine linked to biotin). The code following nnm: can be more descriptive to accommodate the diversity of these modifications, with users expected to employ concise labels. An undefined synthetic modification can be indicated by {nnm} (without a code).

- **Non-natural residues (nnr):** In MAP format, non-natural amino acids are represented directly using the {nnr:ResidueName} tag. For instance, a peptide commencing with norleucine followed by a standard sequence would be represented as {nnr:Nle}GAKT. The tag format {nnr:Code} employs short, descriptive identifiers (e.g., Nle for norleucine, Orn for ornithine, Aib for α-aminoisobutyric acid). For example, the sequence GLY{nnr:Cit}KTR indicates a citrulline residue at the fourth position, and the sequence A{nnr:Dap}CPQ contains diaminopropionic acid at the second position. This notation facilitates the incorporation of a wide array of non-standard residues into peptide and protein sequences, commonly encountered in experimental design, peptide therapeutics, and structural studies. In instances where a user cannot define a specific unnatural residue, it can be represented by the simple {nnr} tag.

- **Conjugation tags (conj):** Proteins and peptides are frequently conjugated to other molecules, forming larger entities (e.g., protein-polymer, protein-lipid, protein-nucleic acid, or drug conjugates). These attachments, beyond simple modifications, involve the addition of substantial molecular moieties. MAP uses {conj:Code} to indicate a residue covalently linked to another molecule or macromolecule (e.g., {conj:Lipid} might mark a glycine undergoing lipidation, such as myristoylation). {conj:Biotin} could indicate biotin conjugation, although this might also be classified as {nnm:Biotin}; the usage can overlap, with conj generally emphasizing larger attachments or the act of conjugation. For example, AC{conj:Mal}GPST signifies a cysteine residue with a maleimide group attached. Conjugation tags facilitate the documentation of site-specific attachments that are not simple PTMs but deliberate coupling of new entities, particularly relevant in drug design and chemical biology.

- **Isotopic/fluorescent labels (iso):** Experimental workflows often involve the incorporation of stable isotopes (e.g., $^{13}C$, $^{15}N$) or the attachment of fluorescent dyes for tracing and imaging. While these do not alter the primary amino acid sequence, annotating their presence is valuable. MAP reserves the iso prefix to cover these labeling cases (e.g., {iso:13C} after a residue indicates $^{13}C$ labeling). Other examples include {iso:15N} for nitrogen-15 labeling and {iso:2H} for deuterium labeling. For fluorescent dyes, shorthand notations or names can be used (e.g., {iso:FITC}, {iso:Cy5}, {iso:Alexa488}). Placing these tags after the labeled residue highlights the site of labeling (e.g., AK{iso:Fluorescein}DGH indicates a lysine conjugated to fluorescein). While many isotopic labeling experiments involve uniform labeling, MAP's site-specific tagging is particularly useful for documenting specific labeling sites or representative sites in experimental designs. The unified iso prefix simplifies the notation for both isotopic and other labels.

All these tag types can coexist in a single sequence, allowing extremely rich annotation.

**Sequence Variants and Mutations**

Proteins frequently exist in multiple isoforms or variants arising from genetic variation, alternative splicing events, or engineered mutations. Traditional sequence formats typically address this heterogeneity by providing multiple distinct sequences or a reference sequence accompanied by a separate list detailing the variations. In contrast, MAP offers a more integrated and compact approach to denote mutations, insertions, and deletions relative to a designated reference sequence, all within a single sequence entry. The fundamental strategy involves treating the provided sequence as the reference (e.g., the wild-type sequence) and

employing specific tags to indicate the nature and location of differences in a variant (See Table 3):

- **Point mutation (mut):** The {mut:X} tag, placed immediately after a residue, indicates that the preceding residue in the reference sequence is replaced by the amino acid denoted by 'X' in a variant. For instance, GPA{mut:R}KV signifies that the amino acid at position 3 in the reference sequence (Alanine, A) is substituted by Arginine (R) in the variant, resulting in the sequence "GPRKV". For clarity, MAP displays the original residue in the sequence and specifies the substituted residue within the tag. In cases involving the replacement of multiple contiguous residues (e.g., a dipeptide replaced by another), the original segment is followed by the {mut:NewSegment} tag. For example, AGRTH{mut:MKN}DGG indicates that the "RTH" segment in the reference sequence is replaced by "MKN" in the variant, yielding a variant sequence of "AGMKNDGG" from the reference "AGRTHDGG".

- **Insertion (ins):** The {ins:SEQ} tag signifies the insertion of the amino acid sequence 'SEQ' (one or more residues) at the position immediately preceding the tag in the variant sequence relative to the reference. The tag is typically placed directly after the residue preceding the insertion site. For example, AG{ins:G}TCA indicates that a Glycine (G) residue is inserted between the G and T of the reference sequence "AGTCA", resulting in the variant sequence "AGGTCA". Similarly, ATRG{ins:MK}TCA denotes the insertion of the dipeptide "MK" after the "ATRG" segment of the reference, producing the variant "ATRGMKTCA". Essentially, the residues flanking the tag in the MAP format represent the context in the reference sequence, while the content within the {ins:...} tag represents the inserted sequence in the variant.

- **Deletion (del):** Deletions are indicated using the {del} tag. For the deletion of a single residue, the {del} tag is placed immediately after that residue in the MAP sequence, signifying its absence in the variant. For example, ARGM{del}KLV implies that the Methionine (M) residue present in the reference sequence is deleted in the variant, resulting in the variant sequence "ARGKLV". For the deletion of a contiguous segment of residues, MAP allows the explicit listing of the deleted segment within the tag, for example, AGH{del:RTH}DGG to indicate the deletion of "RTH". In this case, the reference sequence is "AGHRTHDGG", and the resulting variant is "AGHDGG". While including the deleted segment within the {del:...} tag might appear redundant (as "RTH" also precedes the tag), it enhances clarity regarding the intended deletion. Alternatively, one could place {del} after each residue of the deleted segment or after the entire segment. However, for multi-residue deletions, consistency in notation is crucial, and listing the deleted segment within the tag is recommended to avoid ambiguity.

**Interacting Residues in Proteins**

MAP facilitates the annotation of residues directly involved in molecular interactions through the {IR:Target} tag, where 'Target' specifies the interacting molecule or ligand. This tag is positioned immediately after the interacting residue within the sequence. For example, K{IR:DNA} denotes a lysine residue that participates in DNA interaction, and H{IR:Zn} indicates a histidine residue involved in binding a zinc ion. MAP accommodates a diverse range of interaction targets, including nucleic acids (e.g., DNA, RNA), ions (e.g., Ca, Mg, Zn), small molecules (e.g., ATP, GTP), and macromolecular partners (e.g., proteins, antibodies). These annotations are particularly valuable in structural biology and protein function prediction, as knowledge of interaction sites contributes to understanding binding specificity and molecular mechanisms. In contrast to PTMs or variant annotations, the {IR:} tag describes a functional or structural role of the residue rather than a chemical modification, enabling MAP to encode interaction context directly within the sequence (See Table 3).

**Table 3: Shows variations in sequence and structural annotations in a sequence.**

| MAP Format | Example Sequences | Description of Example |
|---|---|---|
| **Sequence Variants and Mutations** | | |
| {mut:R} | GPA{mut:R}KV | "A" is mutated to Arginine "R" |
| {mut:MKN} | AGRTH{mut:MKN}DGG | "RTH" are replaced by "MKN" |
| {ins:G} | AG{ins:G}TCA | "G" is inserted at 3$^{rd}$ position |
| {ins:MK} | ATRG{ins:MK}TCA | "MK" inserted at 5$^{th}$ position |
| M{del} | ARGM{del}KLV | "M" is deleted from the sequence |
| {del:RTH} | AGHR{del:RTH}DGG | "RTH" are deleted from the sequence |
| **Interacting Residues** | | |
| {IR:DNA} | AT{IR:DNA}IFG | Residue "T" interacts with DNA |
| {IR:RNA} | ADK{IR:RNA}FG | Residue "K" Interacts with RNA |

| {IR:Pro} | ARGG{IR:Pro} | Residue "G" interacts with Protein or peptides |
|---|---|---|
| {IR:ATP} | ACD{IR:ATP}FK | Residue "D" interacts with ATP |
| {IR:GTP} | ACD{IR:GTP}FK | Residue "D" Interacts with GTP |
| {IR:UTP} | ACD{IR:UTP}FK | Interaction with UTP |
| {IR:NAD} | DE{IR:NAD}FG | Interaction with NAD |
| {IR:FAD} | DE{IR:FAD}FG | Interaction with FAD |

**Implementation and Parsing**

To facilitate the adoption and utilization of the MAP format, we developed a prototypical parser capable of both reading and writing MAP entries. This parser interprets any {...} construct encountered within the sequence as an annotation, excluding it from the core amino acid sequence. Furthermore, the parser includes functionality to strip all annotation tags, thereby generating a standard FASTA sequence when required. Given the design principle of maximal simplicity, the implementation of a MAP parser or converter is relatively straightforward. This can be achieved by iterating through the sequence lines character by character, identifying the initiation of an annotation upon encountering a {, and capturing all subsequent characters until the matching }. This captured string is then associated with the preceding residue (or noted as a terminal annotation if located at the end of the line). Similarly, the header line can be processed by splitting the string based on brace characters after the initial identifier. From an implementation perspective, it is crucial that brace characters are exclusively reserved for annotation tags within the sequence. We specifically selected curly braces {} due to their infrequent occurrence in biological sequence data and their absence from the standard set of amino acid symbols. Curly braces offer a visually distinct delimiter that is readily identifiable through regular expressions or dedicated parsing logic.

The MAP format design obviates the need for special escaping mechanisms. We assume that the content enclosed within a {...} tag will not contain a } character (except for the closing brace), and conversely, an { character is not permitted within a tag. Consequently, tag nesting is not supported. This constraint has not presented a limitation in the context of the intended usage, where each tag describes a discrete feature. If multiple features pertain to a single residue, they are represented as a series of consecutive tags following that residue. For instance, a cysteine residue that is both palmitoylated and involved in zinc binding could be represented as C{ptm:Palm}{IR:Zn}.

# Results

## Major Applications

MAP format can be used for many applications, we will describe here its utility in maintaining databases and datasets in a unified format.

## Protein Databases

Numerous protein databases have been developed to manage protein-associated information. UniProt is a highly curated and commonly used resource that maintains comprehensive information about proteins. The MAP format can be effectively applied to enhance UniProt entries by embedding residue-level annotations, such as post-translational modifications (PTMs), non-natural residues, and known mutations, directly within the sequence line. For instance, PTMs, currently detailed in the "Feature" section, could be represented inline using tags like {ptm:} (e.g., phosphorylation, glycosylation), while sequence variants and isoforms could be encoded with {mut:}, {del:}, and {ins:} tags. This integration would allow UniProt to offer MAP as an export option, yielding a unified, annotation-rich sequence format that supports both human readability and machine parsing, without altering the canonical sequence. Protein Data Bank (PDB) is a primary database developed to maintain the tertiary structure of proteins, which is used for structural annotations. The MAP format can be effectively employed to annotate protein and peptide chains derived from PDB structures. Residue-level annotations, including those for post-translational modifications ({ptm:}), cyclizations ({cyc:}), D-amino acids ({d}), non-natural residues ({nnr:}), and molecular interactions ({IR:}), can be extracted from structural data and represented within the MAP format. Furthermore, structural variations and missing segments can be captured using {mut:}, {ins:}, and {del:} tags. This conversion of 3D structural annotations into a compact 1D MAP sequence facilitates downstream sequence-based analyses, alignments, and comparative modelling, while preserving detailed chemical and structural information.

## Database of Peptides

Numerous peptide databases have been developed in the past to curate and maintain peptide-associated information. For example, databases such as IEDB, PRRDB2, MHCBN, AntigenDB, ImmunoSPdb and BCIpep focus on the immunogenic and antigenic properties of peptides and are extensively utilized in vaccine research [26–35]. Currently, these databases employ different conventions for representing peptides capable of eliciting B-cell, T-cell, or antigen-presenting cell responses, resulting in a lack of consistency across platforms. Moreover, the accurate representation of chemically modified peptides remains a significant challenge within these resources. The MAP format may be used to represent these peptides in a unified format across different databases. The increasing number of peptide- and protein-based therapeutics receiving FDA approval over the past two decades has spurred the development of several specialized therapeutic peptide databases. Following CAMPR4, CPPsite2, CancerPDF, PEPlife, B3Pdb, SalivaDB, APD3, CAMP, ViralVacDB, BacVacDB, HMRBASE2, FermFooDb, SATPdb, and DBASSP are a few examples of databases that are used directly or indirectly in the design of peptide-based drugs [13,15,36–46]. Despite the prevalence of chemically modified peptides in these databases, a unified format for their systematic representation is currently lacking. The MAP format addresses this gap by providing a consistent and extensible framework for representing both natural and chemically modified peptides across diverse therapeutic databases

**Bioinformatics Datasets**

The field of protein research has witnessed a significant proliferation of computational methodologies aimed at elucidating various residue-level features and functionalities. For instance, a suite of predictive tools, including DBpred, PPRint2, ATPint, and GTPint, has emerged to identify specific amino acid residues within a protein sequence that are likely to engage in interactions with crucial biomolecules such as DNA, RNA, ATP, and GTP, respectively [47–50]. These methods often rely on machine learning algorithms trained on experimentally validated interaction data. Similarly, the prediction of post-translational modifications (PTMs), which play a critical role in regulating protein function, stability, and localization, has been addressed by tools like GlycoEP and GlycoPP, specifically designed to pinpoint potential glycosylation sites within protein sequences. Furthermore, recognizing the growing importance of chemically modified peptides in therapeutics and materials science, specialized computational methods such as AntiMPmod, CellPPDmod, and HemoPiMod have been developed to predict key properties like antimicrobial activity, cell penetration potential, and hemolytic potential of these engineered molecules [51–53]. These methods often consider the nature and location of the chemical modifications in their predictive models.

A considerable impediment to the seamless integration and comparative analysis of these diverse computational approaches lies in the lack of standardization in the formats used to represent the underlying data and the resulting annotations. Typically, each research group or individual author adopts data formats that best suit their specific analytical pipeline or convenience, leading to a fragmented landscape of incompatible data representations. This heterogeneity complicates efforts to build unified databases, benchmark different prediction methods, or readily share annotated datasets across the scientific community. Another illustrative example of a specialized tool is PepStrMOD, which focuses on predicting the three-dimensional structure of chemically modified peptides [54]. Notably, this method often requires input through a user-friendly web interface, highlighting the need for standardized input formats that can be easily generated and exchanged. The MAP format, with its capacity to encode a rich spectrum of residue-level annotations, including interaction sites, post-translational modifications, non-natural modifications, and even structural features like cyclizations, holds the potential to serve as a unifying framework for representing both the input data and the output annotations generated by these diverse computational methods. By providing a consistent and expressive format, MAP could significantly enhance data interoperability, facilitate the development of integrated bioinformatics platforms, and promote the broader sharing and utilization of valuable protein annotation data.

**Comparison of Different Formats**

The MAP format offers a unique balance between simplicity, expressiveness, and compatibility compared to existing protein and peptide sequence formats. While the FASTA format remains widely used for its minimalism and readability, it lacks support for representing chemical modifications, non-natural residues, or mutations. Formats like HELM and BILN address these limitations by offering detailed structural representations, but their complexity and dependence on predefined monomer libraries hinder widespread adoption. PEFF, an extension of FASTA, supports metadata and variant annotations but stores modification details outside the sequence, reducing immediate interpretability. In contrast, MAP integrates residue-level annotations directly within the sequence using in-line curly-brace tags, while simultaneously encoding protein-level metadata in the header. This dual-layer structure allows MAP to represent complex biological and synthetic sequences in a compact, human-readable, and tool-friendly format. Its backward compatibility with FASTA and flexibility in encoding diverse features make it well-suited for both computational analysis and database integration. The following is a detailed comparison with other formats.

**Comparison to FASTA:** MAP is essentially an extension of FASTA, and thus it inherits FASTA's strengths (simple structure, one-letter codes) while overcoming some key weaknesses. Standard FASTA files cannot natively represent modifications or non-standard residues; any such information must be conveyed externally (e.g., in database comments). In contrast, MAP embeds this information directly. From a user perspective, MAP entries look like familiar FASTA entries but with extra informative text. The design choice to ensure that removing all braces yields a valid FASTA sequence means that MAP is backward-compatible in principle. Overall, compared to plain FASTA, MAP offers richer information density at the cost of a slightly more complex syntax. We believe this trade-off is justified for use cases where modifications matter, and it can be managed such that any pipeline not interested in modifications can treat MAP entries as if they were FASTA after a quick clean-up.

**Comparison to HELM:** The HELM notation marked a substantial advancement in the representation of biopolymers by accommodating non-linear architectures, such as branched and cyclic peptides, as well as conjugates with diverse entities. This capability contrasts with the fundamentally linear nature of MAP, which lacks the inherent capacity for explicit representation of such complexities. For instance, a peptide comprising two crosslinked chains is naturally encoded in HELM as two distinct polymer strands connected via a defined linkage. While MAP could attempt to approximate this scenario through concatenation of the two sequences with a specialized delimiter and the application of {cyc:X-Y} tags to denote the crosslink, this approach lacks elegance and clarity. Consequently, MAP is not intended as a replacement for HELM in instances of highly intricate topologies. Instead, MAP is designed for scenarios where a single linear sequence, augmented with annotations, provides a reasonable and effective representation. Furthermore, HELM necessitates referencing a predefined monomer library for non-standard residues, whereas MAP offers the flexibility of incorporating descriptive tags dynamically. A key advantage of MAP in routine laboratory practice lies in its inherent readability and ease of integration; a bench scientist can likely interpret a MAP sequence intuitively, whereas HELM typically requires specialized training or decoding tools for comprehension.

**Comparison to BILN:** The BILN format shares a philosophical alignment with MAP in its pursuit of human readability and its capacity to represent complex peptides. BILN employs a notation that explicitly encodes multiple chains and cyclic structures through the use of identifiers and indices, enabling precise specification of bond formation involving specific amino acid R-groups. This design confers significant expressivity upon BILN, allowing for the accurate definition of cyclic architectures, branching points, and even non-peptide attachments, contingent upon monomer definition. However, the inherent complexity of handling these features necessitates the introduction of additional symbols and grammatical elements within BILN, such as parentheses and numerical identifiers. Furthermore, BILN was primarily conceived for the accurate chemical depiction of peptides, with an emphasis on facilitating structure generation via tools like RDKit. In contrast, MAP maintains a format more closely resembling a conventional sequence, interspersed with annotation tags. One could characterize BILN as more chemistry-centric, while MAP exhibits a stronger sequence-oriented focus.

**Comparison to PEFF:** The PSI Extended FASTA Format (PEFF) represents another strategy for incorporating modifications and variants into sequence data by appending specific key-value pairs within the FASTA header or as comment lines. For example, PEFF might denote a variant as VariantSimple=123 A T, indicating an alanine to threonine substitution at position 123, or list post-translational modifications (PTMs) using a controlled vocabulary. The design principle ensures a degree of backward compatibility, such that legacy tools encountering a PEFF file would either interpret the added information as comments or as specially formatted headers that could be safely ignored, thus minimizing disruption. A key distinction between PEFF and MAP lies in their approach to annotation. In PEFF, the sequence line adheres strictly to standard amino acid representations, with all annotation data relegated to metadata fields. Conversely, MAP integrates annotations directly within the sequence line. The primary advantage of PEFF's approach is the preservation of an unmodified sequence line, ensuring that any tool parsing the sequence will retrieve a completely standard amino acid string. However, this separation of annotation from the corresponding residues can render the information less immediately intuitive and necessitates multi-line records, potentially complicating tasks such as copy-pasting. MAP's inline tags, by contrast, provide immediate contextual information, placing the modification or variant directly adjacent to the affected residue. This proximity can facilitate the understanding of sequence-function relationships, such as the identification of clusters of modified residues. Both PEFF and MAP share the foundational concept of building upon the established FASTA format and prioritizing flexibility in representation. Indeed, these formats could be viewed as complementary approaches, with a potential future integration where MAP-like inline tags serve as one option for encoding features within a PEFF specification. Currently, however, they have emerged from distinct communities, with PEFF originating from proteomics standards initiatives and MAP arising from a need for a user-friendly format. For the typical biologist or bioinformatician, MAP may offer greater ease in manual editing and interpretation, while PEFF's rigid schema might be more advantageous for large-scale computational pipelines designed to parse specific, well-defined fields.

**Repository for MAP Format**

To facilitate the scientific community's adoption of the MAP format in their research endeavours, we have developed a comprehensive web-based portal, MAPrepo, accessible at https://webs.iiitd.edu.in/raghava/maprepo/. This portal offers a suite of resources designed to support users, encompassing the following key modules:

- **Manual:** This module provides comprehensive documentation on the MAP format, including detailed specifications, illustrative examples, and insightful case studies.
- **Databases:** This section catalogs existing biological databases that have already integrated the MAP format for representing peptide and protein sequences, thereby promoting interoperability and data sharing.
- **Datasets:** Recognizing the importance of standardized data for method development, this module presents a collection of relevant datasets curated in the MAP format.
- **Python Scripts:** We are actively developing a collection of Python scripts to streamline the conversion, management, and maintenance of data in MAP and other common bioinformatics formats. Notably, this includes scripts for the conversion of Protein Data Bank (PDB) format files into MAP format.
- **Web Apps:** This module hosts user-friendly web applications designed to facilitate the practical application of the MAP format in research workflows.

Through MAPrepo, we aim to provide a centralized and accessible resource that empowers researchers to effectively utilize the MAP format in their investigations, fostering enhanced data representation and analysis within the bioinformatics community.

**Discussion**

The introduction of the MAP format addresses a salient need within bioinformatics for a sequence representation that transcends the limitations of a bare amino acid string without compromising usability. MAP has many advantages over existing alternative formats, following is brief description. MAP uniquely integrates sequence and annotation within a unified framework. MAP facilitates direct manual annotation of sequences without the prerequisite of specialized software, yielding readily interpretable results. The adoption of plain text and straightforward delimiters ensures compatibility with standard text editors. As an extension of the widely adopted FASTA format, MAP allows for seamless integration into existing bioinformatics pipelines with minimal modifications. In scenarios requiring strict FASTA compliance, tags can be computationally and trivially removed using regular expressions or simple scripts. MAP exhibits broad applicability, encompassing a wide spectrum of use cases, ranging from naturally occurring post-translational modifications and genetic variants to synthetic modifications and conjugations. By providing a richer and more informative data representation, MAP serves as a catalyst for the development of novel bioinformatics tools and the enhancement of existing ones. For instance, multiple sequence alignment viewers could leverage MAP to highlight conserved PTM sites across orthologous sequences, or proteomics database search engines could directly parse MAP format, enabling the explicit consideration of dynamic modifications at specified residues, thereby refining search parameters by distinguishing between fixed and variable modifications. It can also be used to incorporate regular or irregular secondary structure of proteins like helix, turns, coils, beta-sheets etc. [55–59].

Despite its considerable utility, MAP have a number of limitations. Firstly, its fundamental nature as a linear sequence format restricts its applicability. Proteins that cannot be reasonably represented as a linear amino acid sequence, even with the inclusion of loop-indicating tags, are not well-suited for MAP. Examples include proteins with multiple disulfide-linked chains that are not contiguous in sequence, such as insulin with its separate A and B chains. While one could artificially concatenate these chains in MAP and denote the disulfide bond, this representation lacks intuitive clarity. In such cases, a multi-entry approach or an alternative format may be more appropriate. Secondly, the current MAP specification largely lacks enforced controlled vocabularies for many annotation tags, with the exception of the standardized tags we have defined. This absence of strict standardization could lead to inconsistencies in annotation; for example, phosphorylation might be annotated as {ptm:Phospho} by one user and {ptm:Phos} by another. Over time, community consensus should facilitate the emergence of best practices, and community guidelines could potentially refine these vocabularies. This evolutionary process mirrors the historical convergence on modification notations in PDB files and the development of controlled vocabularies like PSI-MOD, which could be readily mapped to MAP tags. Thirdly, the inclusion of tags inherently increases the length of sequence strings, potentially impacting analyses such as sequence identity calculations or motif searches if not appropriately addressed.

For MAP to achieve broad utility, its adoption or at least recognition by major bioinformatics platforms is crucial. Integrating MAP support into widely used libraries such as BioPython and BioPerl would significantly facilitate its uptake within the community. We are committed to providing reference implementations and actively encourage such integration efforts. Envisioning a future where databases like UniProt offer MAP format exports of their entries, incorporating modifications and variants inline, would greatly streamline data retrieval for researchers interested in these features. Furthermore, sequence alignment programs could offer the option to output alignments in MAP format, visually highlighting conserved modifications. MAP's line-based structure and relatively simple syntax (eschewing complex parsers like those required for JSON or XML) also render it highly amenable to web applications, enabling user-friendly interfaces for visual sequence annotation and subsequent MAP format export using web programming languages like JavaScript. The current MAP specification primarily focuses on proteins and peptides. However, the underlying concept of in-sequence tags could, in principle, be extended to nucleotide sequences to denote modified nucleotides, DNA methylation, RNA modifications such as pseudouridine, and other relevant features. A distinct set of prefixes (e.g., {mod:5mC} for 5-methylcytosine) could be employed for this purpose. Nevertheless, such an expansion falls outside our current scope and would necessitate careful consideration to avoid conflicts with existing nucleotide ambiguity codes. Another potential extension involves enabling the mapping of modifications to specific atoms or sub-residues when a higher level of detail is required, such as distinguishing phosphorylation on a particular hydroxyl group within a multi-hydroxyl amino acid. However, this level of granularity leans towards detailed chemical representation and may exceed the requirements of MAP's intended applications.

As with any proposed standard, community feedback will be indispensable for the refinement and evolution of MAP. We anticipate that as users apply MAP to diverse and challenging edge cases, they will provide valuable suggestions for new tags or improvements to the existing specification. For instance, users might propose shorthand notations for representing common scenarios such as disulfide bond patterns across multiple cysteines or more concise ways to indicate sequence repeats or variants. We actively welcome such input and view the current MAP specification as a robust but adaptable framework capable of evolving based on community needs. The core principles of human readability and base FASTA compatibility will remain central to any future modifications.

In conclusion, MAP represents a practical and user-centric format that effectively bridges the gap between raw sequence data and detailed chemical annotations. While it does not aim to supplant formats designed for comprehensive chemical structure representation, it significantly enhances the annotation capabilities of everyday protein sequence formats. By comparing MAP with existing formats such as FASTA, HELM, and BILN, we observe that MAP occupies a unique and valuable niche. It offers a greater capacity for annotation than FASTA while maintaining a level of simplicity and sequence-centricity that distinguishes it from the more chemically oriented HELM and BILN formats. This positions MAP as an attractive and practical option for a wide range of applications in proteomics, genomics (particularly for coding sequence variations), and any context where the integration of protein identity and its associated modifications is paramount.

## Conflict of interest

The authors declare no competing financial and non-financial interests.

## Authors' contributions

GPSR and AS implemented the idea. NKM, NK, AS, SP, and ASR create the scripts. ASR created the front-end, back-end of the web server. AS, NKM, NK, SP, ASR, and GPSR penned the manuscript. GPSR conceived and coordinated the project. All authors have read and approved the final manuscript.

## Acknowledgments

Authors are thankful to the University Grants Commission (UGC), Department of Biotechnology (DBT), DBT-RA program in Biotechnology and Life Sciences for fellowships and financial support, and the Department of Computational Biology, IIITD New Delhi, for infrastructure and facilities. We would like to acknowledge that the Figures were created using BioRender.com.